\documentclass[lettersize,journal]{IEEEtran}
\usepackage{tikz}
\usepackage{amsmath}

\usepackage{filecontents}

\usepackage{cite}
\usepackage{amsmath,amssymb,amsfonts}
\usepackage{algorithmic}
\usepackage{graphicx}
\usepackage{textcomp}
\usepackage{xcolor}
\usepackage{url}
\usepackage{hyperref}
\usepackage{multirow}
\usepackage{comment}
\usepackage{ulem}
\usepackage{subcaption}

\usepackage{array}

\def\BibTeX{{\rm B\kern-.05em{\sc i\kern-.025em b}\kern-.08em
    T\kern-.1667em\lower.7ex\hbox{E}\kern-.125emX}}

\usepackage{algorithm}
\usepackage{stfloats}
\usepackage{verbatim}
\usepackage{lscape}
\usepackage{soul}

\begin{document}

\title{Impedance vs. Power Side-channel Vulnerabilities: A Comparative Study}

\author{\IEEEauthorblockN{Md Sadik Awal, Buddhipriya Gayanath, and Md Tauhidur Rahman}\\
\IEEEauthorblockA{\textit{Security, Reliability, Low-power, and Privacy (SeRLoP) Research Lab \\ECE Department, Florida International University, Miami, Florida, USA} \\
E-mail: \{mawal003, bgaya003 and mdtrahma\}@fiu.edu}
}

\maketitle

\begin{abstract}
Physical side channels emerge from the relation between internal computation or data with observable physical parameters of a chip. Previous works mostly focus on properties related to current consumption such as power consumption. The fundamental property behind current consumption occur from the impedance of the chip. Contemporary works have stared using chip impedance as a physical side channel in extracting sensitive information from computing systems. It leverages variations in intrinsic impedance of a chip across different logic states. However, there has been a lack of comparative studies. In this study, we conduct a comparative analysis of the impedance side channel, which has been limitedly explored, and the well-established power side channel. Through experimental evaluation, we investigate the efficacy of these side channels in extracting stored advanced encryption standard (AES) cryptographic key on a memory and analyze their performance. Our findings indicate that impedance analysis demonstrates a higher potential for cryptographic key extraction compared to power side-channel analysis (SCA). Moreover, we identify scenarios where power SCA does not yield satisfactory results, whereas impedance analysis proves to be more robust and effective. This work not only underscores the significance of impedance SCA in enhancing cryptographic security but also emphasizes the necessity for a deeper understanding of its mechanisms and implications. 
\end{abstract}


\section{Introduction} \noindent
In cybersecurity, side-channel attacks have long been recognized as potent threats, exploiting unintended information leakage to compromise the security of cryptographic systems and electronic devices. Traditional side channels, such as power,  electromagnetic (EM) emissions, and timing variations, have received significant attention in academia and industry due to their ability to leak sensitive information from cryptographic devices and systems \cite{power_survey,gattu2020power,del2015side,das2020killing,sehatbakhsh2020emsim,agrawal2003side}. For instance, power side-channel attacks leverage fluctuations in the power consumption of cryptographic devices across different operations, such as encryption and decryption processes. Similarly, EM emissions arise from the electrical operations within cryptographic devices, leading to unintentional radiations that attackers can intercept and analyze \cite{das2020killing,sehatbakhsh2020emsim}. Usually, the vulnerabilities in the hardware implementation or the physical characteristics of the encryption systems facilitate these unintended physical side channels, which are then exploited to extract confidential information from cryptographically secure algorithms \cite{power_survey,gattu2020power,del2015side,das2020killing,sehatbakhsh2020emsim,agrawal2003side}.

Recently, impedance analysis has found applications across various domains, including radio frequency (RF) related fields, counterfeit chip detection and tempering detection~\cite{impedance_in_rf,counterfeit_impedance,awal2022nearfield}. 
It has also been discovered that impedance can inadvertently leak sensitive information related to the computations and data being processed within a system.  
Our research demonstrates the potential of the runtime impedance of a device as a side-channel, unveiling software instructions through distinct impedance profiles generated by each operation~\cite{awal2023impedance,awal2022utilization}. Subsequent studies have exploited impedance as a side channel to extract the AES key from the register of an FPGA~\cite{leakyohm}. These approaches to leveraging impedance marks a significant shift in the perception of impedance within the cybersecurity domain. Impedance analysis, while it remains an important tool for enhancing hardware security, it also presents a novel vulnerability that adversaries can exploit to their advantage.

Despite the extensive focus on current and voltage as the primary sources of side-channel leakage, impedance, the parameter that connects these two, offers a distinct and underexplored perspective for SCA. Unlike power-based techniques, which primarily measure current fluctuations, impedance captures the dynamic interaction between current and voltage, providing a distinct and potentially more insightful view into data-dependent variations within a system. 
Although impedance has indirectly contributed to static power analysis attacks, its direct significance has often been overlooked~\cite{10679507,moos2019static,kabin2024exploiting}. 
This is primarily due to the conventional assumption that impedance remains constant, determined by the chip's physical structure, dimensions, and fabrication materials. However, the influence of operational state changes on impedance, and their potential contribution to side-channel leakage, has not been comprehensively investigated. 
This paper addresses this gap by investigating the differences between impedance and power side channels, with a focus on demonstrating the effectiveness of impedance in cryptographic key extraction. Through a comprehensive analysis, this study aims to highlight the potential of impedance as a side-channel vector, offering a new perspective on hardware vulnerability assessment. 
The primary contributions of this paper are, 
\begin{itemize}
  \item Investigating the sensitivities of impedance and power in SCA. Although these side channels are interconnected, this research clarifies how they can exhibit differing sensitivities.  
  \item Exploring impedance side-channel leakage sources and  AES key extraction using this method, including scenarios with and without the presence of background noise.  
  \item Performing a comparative analysis of the performance of impedance and power side channels. 
\end{itemize}

The structure of this paper is organized as follows:  Section \ref{sec:background} discusses the background of power and impedance side channels, including a discussion on correlation analysis and the existing research on the extraction of encryption keys using physical side channels.  
Section \ref{sec:method} presents the proposed method to compare the power and impedance side channels. 
Section \ref{sec:experimental_setup}  details the experimental setup. We present and compare the findings of AES key extraction using power and impedance side channels in Section \ref{sec:results}. Section \ref{sec:conclusion} concludes the work.

\section{Background and Related Work}
\label{sec:background} \noindent
This section provides an overview of the cryptographic algorithm targeted in this study and the concepts of power and impedance side channels. It then outlines the attack methodology used in our analysis. Additionally, recent literature on power and impedance side-channel research is reviewed to contextualize the existing work and highlight the contributions of this study. 

\subsection{Advanced Encryption Standard}
\label{sec:AES} 
\begin{figure}[!htbp]
\centering
\includegraphics[width=0.33\textwidth]{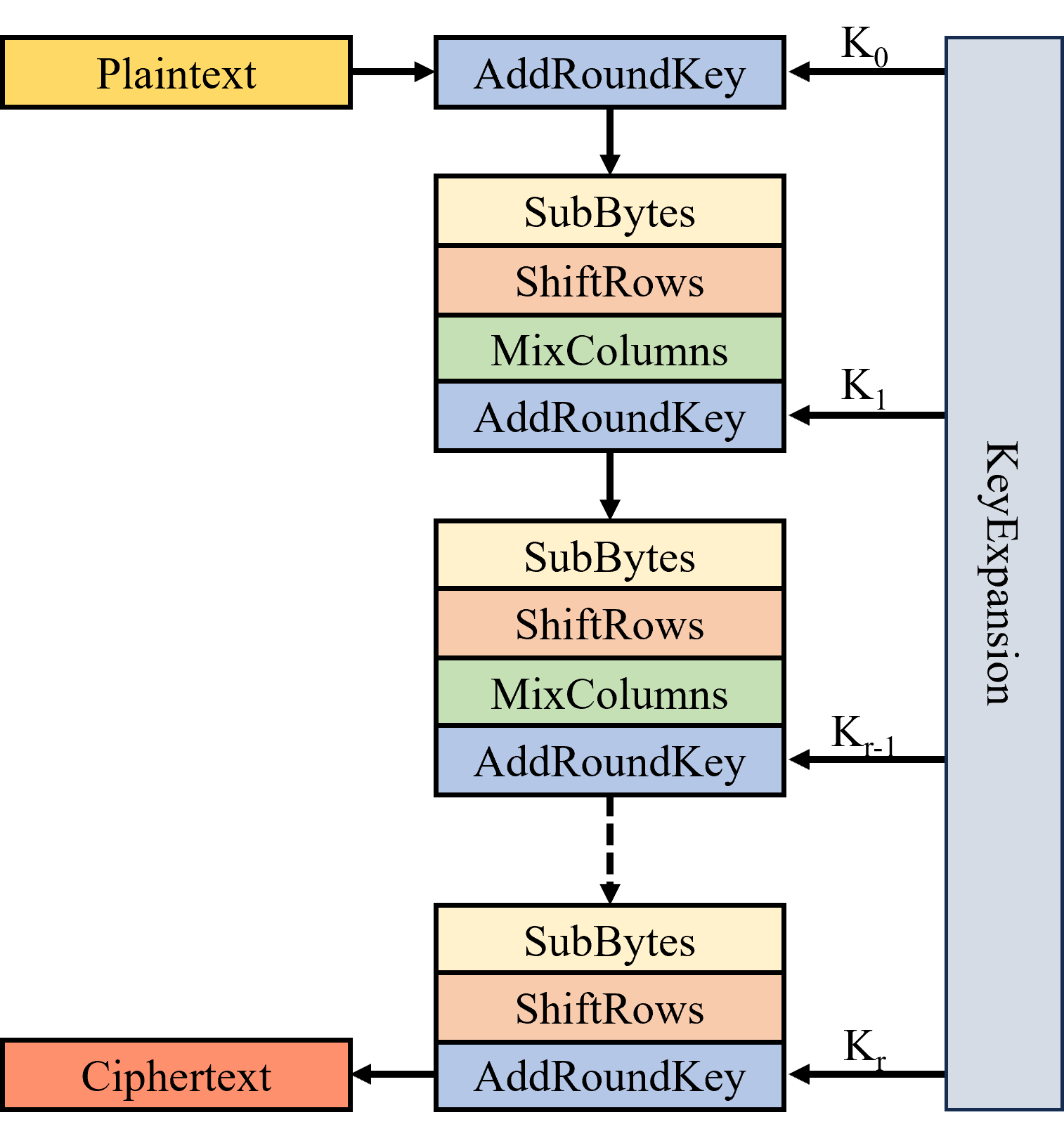}
\caption{Overview of AES Encryption.}
\label{fig:AES_imp}
\end{figure}

\noindent
This study focuses on the AES, a widely adopted  symmetric block cipher that encrypts and decrypts data with the same key. Fig.~\ref{fig:AES_imp} presents the AES encryption process, which consists of $r$ rounds. Each round includes key-dependent operations: substitution using the Rijndael S-BOX~\cite{satoh2001compact,barrera2020fast},  row shifting, column mixing, and round key addition~\cite{NIST}. 
The Rijndael S-BOX, implemented as a lookup table, plays a critical role in security by transforming an 8-bit input into an 8-bit output, ensuring that small changes in input lead to significant alterations in the output. 
In this work, the AES-128 algorithm is used to compare the effectiveness of power and impedance side channels in cryptographic analysis.

\subsection{Power Side Channel}
\label{sec:power}  \noindent
The power side channel exploits the power consumption patterns of target devices and is a well-established area of research~\cite{power_survey}. In digital circuits, power consumption is directly related to the switching activities of transistors during computation. These switching events cause fluctuations in current, leading to measurable variations in power consumption. By capturing voltage and current changes at the power supply pins of the target device using an oscilloscope, these variations can be recorded for further analysis. Such measurements provide indirect information about the executed operations, potentially exposing sensitive data like cryptographic keys, which poses significant security and privacy risks.

This work focuses on exploiting power side channels to extract the AES encryption key. We target an embedded system running AES. We investigate the dynamic power consumption of internal components, such as memory modules, where intermediate values are temporarily stored during the encryption process. The power consumption during memory read and write operations is inherently linked to the switching activities of transistors as they transition between ON and OFF states. 
However, rather than directly measuring these transitions, power side channels capture the cumulative impact of switching activities on the overall power consumption. By analyzing these patterns, particularly during write operations when state transitions occur, we aim to recover the AES encryption key, highlighting the vulnerability of embedded systems to power-based side-channel attacks. 

\subsection{Impedance Side Channel}
\label{sec:impedance} \noindent
In contrast to power side channel, impedance side channel exploits variations in a system's intrinsic impedance, which fluctuates with different logic states. These states correspond to various configurations of transistors, leading to measurable impedance changes~\cite{awal2023disassembling,awal2023impedance}. Such variations are particularly pronounced in memory cells, where distinct impedance differences correspond to the binary data stored.  
This effect is evident across clock cycles in edge-triggered synchronous systems, as the number of transistors in ON and OFF states remains constant until the next clock signal. By exploiting these impedance variations, we focus on analyzing the storage of intermediate values during AES encryption in memory chips. This approach enables the extraction of cryptographic keys through impedance measurements. 

\begin{equation}
Z_{i} = \frac{1+S_{11}}{1-S_{11}} \times Z_{ref}
\label{eq:S_to_Z}
\end{equation}
To measure the impedance, we use a vector network analyzer (VNA). VNA is widely used to characterize multi-port networks. The VNA operates by injecting a known test signal into the device under test (DUT) and analyzing the reflected and transmitted signals. Through this process, the VNA calculates the DUT's scattering parameters (S-parameters), which are important to RF and microwave system analysis. Our analysis focuses on the input port reflection coefficient, $S_{11}$, from which we derive the supply side impedance of the DUT, $Z_{i}$, using Eq.~\ref{eq:S_to_Z}, where $Z_{ref}$ denotes the reference impedance. These impedance measurements form the basis of our work, allowing us to evaluate  the feasibility of using impedance side channels for cryptographic key extraction.

\subsection{Correlation Analysis} 
\label{sec:correlation}  \noindent
Recent works use various analytical methods to study signals in physical side channels. One popular analysis method is correlation-based analysis. This analysis is considered powerful in unmasking the hidden relationships between secret information and noisy measurements using the Pearson correlation coefficient as a statistical test~\cite{CPA_1}. Additionally, rather than relying on a single measurement point, correlation-based analysis uses a multivariate approach by analyzing multiple points within the measurement trace, thereby improving the robustness and accuracy of the side-channel attack~\cite{power_survey}. 

\begin{align}
\rho{(x_{s},h_{k})} &= \frac{M \sum_{m=1}^{M} x_{s,m} h_{k,m} - \sum_{m=1}^{M} x_{s,m} \sum_{m=1}^{M} h_{k,m}}{\sqrt{D_x D_h}} \label{eq:pearson} \\
D_x &= M\sum_{m=1}^{M} x_{s,m}^{2} - (\sum_{m=1}^{M} x_{s,m})^{2} \nonumber \\
D_h &= M\sum_{m=1}^{M} h_{k,m}^{2} - (\sum_{m=1}^{M} h_{k,m})^{2} \nonumber
\end{align}

Correlation-based analysis begins  with the collection of leakage responses generated during known plaintext operations.  Each measurement is synchronized with the cryptographic encryption operation performed in the device over the plaintext. 
The next step involves selecting a leakage model that represents hypothesized side-channel leakages. This leakage model is then correlated with the actual observed leakage to find the leaked information. 
To facilitate this analysis, let $M$ represent the total number of measurements, $S$ the total number of sampling points per measurement, $x_{s,m}$ the actual leakage at the $s$'th sample point of the $m$'th measurement, and $h_{k,m}$ the hypothesized leakage for a guessed key $k$ at the $m$'th measurement. The relationship between the actual and hypothesized leakages is quantified by the Pearson correlation coefficient $\rho{(x_{s},h_{k})}$, calculated as presented in Eq.~\ref{eq:pearson}.

The correct key from the leakage is identified by finding the guessed key for which the maximum absolute correlation occurs within the full sampling point $S$. We use this correlation-based analysis method for both the power and impedance side channels to extract the cryptographic key.

\subsection{Related Work}
\label{sec:relatedwork} \noindent
Power SCA on both the hardware and software implementations of cryptographic algorithms has been extensively researched over the past few decades. 
Authors in \cite{SPA} present one of the earliest power side-channel attacks to extract the AES key from smart cards. In \cite{DPA_anatomy}, the authors discuss a detailed explanation of step-wise differential power analysis (DPA) attacks over AES implementation. 
In \cite{DPA_attack_2008}, three attack scenarios are presented: single-bit DPA, multi-bit DPA, and a correlation analysis. The authors implement AES in two platforms. The study concludes by proposing an enhanced DPA approach that strategically organizes plaintext inputs to maximize the leakage. 
Additionally, authors in \cite{MC_1, MC_2} present AES key extraction attacks using power analysis on the ATmega328P microcontroller. Further, the authors in \cite{AES_ASIC} present a power analysis attack specifically targeted against ASIC implementations of the AES. 
The authors in \cite{advanced_modes_AES,advanced_modes_AES_2} present their findings on power side-channel attacks targeting the advanced modes of AES. Additional works, such as \cite{kabin2024exploiting, moos2019static, 10679507}, use the static power side channel by analyzing the static power consumption of a device. While impedance is a contributing factor in static power consumption, these studies focus exclusively on the effects of impedance changes as they relate to power consumption. 

While impedance has always been a factor in SCA, it has only recently gathered attention in the realm of hardware security~\cite{backscattering,okamoto2012identification,leakyohm}. Prior to its recognition for cryptographic key recovery, many studies focused primarily on other factors. For example, the authors in \cite{zhu2023pdnpulse} introduce a detection technique using board level impedance to detect physical modifications of the board. The study in \cite{backscattering} explores a method for non-destructively detecting hardware Trojans through internal impedance variations. The authors in \cite{backscattering_2}, study a clustering method to classify ICs into Trojan-free and Trojan-infected groups.  
In addition to the usage of impedance in physical modification detection, it is recently attracting the focus of device switching activities. Authors in \cite{awal2023disassembling} use impedance measurements to disassemble software instruction types for anomaly monitoring and software integrity verification. The authors in \cite{cheng2020digital} use the impedance of digital circuits to introduce a novel RFID tag design. In \cite{awal2022utilization}, the study of impedance changes induced by the switching activities is used to detect and identify the firmware. Authors in \cite{leakyohm} use the impedance side channel to extract the AES key from the registers of an FPGA. 
These works present the run-time impedance variations of the devices as the side-channel. 

However, to the best of our knowledge, no unified setup has been established to compare the sensitivities of impedance and power side channels. Therefore, this work primarily focuses on studying this sensitivity.  


\section{Power vs Impedance: Key Extraction} 
\label{sec:method} \noindent
In this section, we discuss the attack methodology used against the AES implementation, focusing on the exploitation of both power and impedance side channels. We detail the threat model, followed by the leakage models used in the attack. Additionally, we discuss the metrics used to evaluate the effectiveness of both power and impedance side channels.

\subsection{Source of Impedance Side-channel} \noindent
Impedance SCA introduces a novel approach to probing the intricate and often overlooked behaviors of chips, particularly in memory chips. This technique methodically explores memory operations, revealing how variations in impedance could potentially disclose information about the data being processed and stored. Central to this analysis is the understanding that multiple physical and architectural factors influence the impedance characteristics of a chip~\cite{awal2023disassembling,backscattering}.These factors include parasitic elements inherent to the physical design, interconnection structures, variations in transistor properties across the silicon substrate, and the distinct semiconductor material characteristics associated with different transistor configurations. 

In memory chips such as static random access memory (SRAM) and dynamic random access memory (DRAM), the control circuit manages memory access and data read/write operations. The memory array, organized in rows and columns, stores data at specific address locations. Three primary components define the impedance profile of memory: (i) the architecture and layout of the memory array, (ii) the cell type, and (iii) the associated  data and control logic circuitry. Manufacturers tailor the arrangement of memory cells in rows and columns, as well as select the cell type (e.g., four-transistor (4T) or six-transistor (6T) cell structures in SRAM), based on factors such as performance requirements, power consumption, and area constraints, all of which influence the impedance characteristics. 

In our study, we specifically focus on the impedance changes that correspond to the content stored in memory cells. The process of writing data to or reading from memory cells initiates a series of events that create unique impedance profiles for the bit sequences stored within the memory array. Factors such as transistor activation, variations in the fabrication process, and specific data retention logic used contribute to these distinctive impedance profiles. Furthermore, the profiles are further influenced by the nature of the data stored in the cells, their locations within the array, and the varying lengths of interconnections for each cell. Thus, impedance measurements are expected to reflect these variations, making them susceptible to information leakage.

\subsection{Attack against AES} \noindent
In this subsection, we describe the target intermediate of AES encryption, as well as the hypothetical leakage model.  

\subsubsection{Target Intermediate}
The key extraction of the AES encryption using physical SCA usually targets the first or final round of the algorithm. The targeted round is used to model the leakage signal. In our study, we perform impedance SCA on the first round of the AES algorithm, targeting the output of the S-Box. The process begins with a bitwise XOR operation between the plaintext and the secret key, followed by the substitution step using the S-Box. The output value of the substitution step is stored in memory, and it is our targeted computational intermediate value. The steps of AES encryption are illustrated in Fig.~\ref{fig:AES_imp}.

In our hypothetical leakage model, we use the plaintext $p_i$, and the corresponding round subkey $k_j$ to define the targeted computational intermediate $m_i$, as presented in Eq.~\ref{eq:sbox}. 
This intermediate value is influenced by both the input plaintext and the secret cryptographic key, establishing a correlation between the side-channel leakages and the secret cryptographic key. Attackers can exploit the statistical properties of the leakage traces to extract the cryptographic key. 
By focusing on the first round and the S-Box output while effectively modeling the leakage signal, we aim to enhance the understanding of how impedance SCA can compromise cryptographic security.  

\begin{equation}
m_{i} = SBOX(p_i \oplus k_j)
\label{eq:sbox}
\end{equation}

\subsubsection{Leakage Model}
We perform correlation analysis-based attacks against the AES implementation using both power and impedance side channels and evaluate their results.
We use the hamming weight (HW) leakage model \cite{peeters2007power,kumar2022side,randolph2020power}. The HW refers to the total number of set bits (number of `1's) in the data. This HW leakage model is predicated on the assumption that the power consumed by the cryptographic device—or the impedance variations~\cite{peeters2007power,randolph2020power} observed—is directly correlated with the HW of the processed data. As the hamming weight increases, more transistors switch, leading to higher leakage.  

The HW leakage model considers the HW of the targeted intermediate value as the leakage source. It assumes a linear relationship between the HW and the observed leakage. Let $L$ be the observed leakage, $HW(m)$ be the hamming weight of the targeted intermediate $m$, $a$ and $b$ be the constant coefficients, and $n$ be the noise in the leakage measurement. The observed leakage, $L$, can be expressed as in Eq.~\ref{eq:HW}. 
The fundamental assumption here is that variations in the physical properties observed through side channels during cryptographic operations can reveal information about the internal data being processed. 
\begin{equation}
L = a \times HW(m_i) + b + n
\label{eq:HW}
\end{equation}

The effectiveness of the HW model in side-channel attacks lies in its ability to exploit the statistical correlation between the measured leakages and the HW of the secret data being processed. We can make informed guesses about the secret key by analyzing these correlations.


\subsection{Threat Model} \noindent
During the experiment, we write the intermediate values to the memory. 
It is assumed that the attacker can control the input plaintexts to the system. The assumptions for the experiment are as follows.
\begin{itemize}
    \item In the power SCA, the attacker can access the power delivery network (PDN) of the memory chip and measure the power signals when the data is written to the memory using an oscilloscope. 
    \item In the impedance SCA, the attacker can connect a VNA to the PDN of the memory module and measure the impedance after each intermediate value corresponding to a plaintext is written to the memory.
\end{itemize}

\subsection{Evaluation Metric} 
\label{sec:metric} \noindent
To compare the performance of power and impedance side channels, we employ three comprehensive metrics. These metrics do a systematic analysis of the effectiveness of both power and impedance side channels in extracting the AES key.  

In the evaluation of the two targeted side channels, we use correlation ratio (CR) Eq.~\ref{eq:corr_ratio_modified} as a confidence metric for correct key recovery. It is calculated as the ratio between the correlation coefficient for the correct key and the highest correlation coefficient observed for the wrong keys. It can assess the confidence of the correct key guess with respect to other wrong key guesses. A higher ratio indicates a strong level of confidence in precisely guessing the correct value for the key.  Let $k_{c}$ and $k_{w}$ denote the correct subkey and the wrong guessed subkey, respectively, from $K$ key space. We use Eq.~\ref{eq:pearson} to develop Eq.~\ref{eq:corr_ratio_modified} and compute the CR as of the confidence score of each subkey. 

\begin{equation}
CR = \frac{ \max_{m \in M} (\rho(x_{m}, k_{c})) }{ \max_{m \in M} (\rho(x_{m}, k_{w})) \text{ for } k_{w} \in K }
\label{eq:corr_ratio_modified}
\end{equation}

As for the second metric, we use the minimum traces to key disclosure (MTD) metric to evaluate which side channel will result in early identification of the correct key with respect to the minimum number of traces. The significance of MTD lies in its ability to quantify the data complexity of side-channel attacks, assessing the practical feasibility and security implications of such attacks. 
A lower MTD value indicates more efficient and effective side-channel attacks, as it requires fewer traces to extract the secret information. On the other hand, a higher MTD value suggests a less efficient side channel, requiring more traces to extract the subkey and, consequently, may be less practical in real-world scenarios. Thus, MTD can be used to compare the efficacy of the power and impedance side-channel attacks. 
 

In the evaluation for assessing the effectiveness of side-channel attacks, we use the interquartile range (IQR) as our third metric. The IQR serves as a robust measure for outlier detection within datasets, which is essential for ensuring the integrity of the data under analysis. 
The IQR, as defined in Eq.~\ref{eq:iqr}, calculates the difference between the 75th percentile (Q3) and the 25th percentile (Q1) of the observed data points. This calculation focuses on the variability within the central 50\% of the data and renders the IQR a pivotal tool in identifying significant deviations in the correlation coefficients calculated using the hypothetical keys to find the correct key. 

\begin{equation}
\text{IQR} = Q3 - Q1
\label{eq:iqr}
\end{equation}

\begin{equation}
\text{Outliers} =
\begin{cases}
x < Q1 - 1.5 \times \text{IQR} \\
x > Q3 + 1.5 \times \text{IQR}
\end{cases}
\label{eq:iqr_outliers}
\end{equation}

The outliers of the correlation coefficient distribution formed by all guessed keys represent the potential correct key candidates. Let $x$ be the calculated correlation coefficient using Eq.~\ref{eq:pearson} due to the guessed subkey $K_{x}$. By establishing lower and upper bounds for outlier detection, a correlation coefficient $x$ outside these thresholds can be flagged as an outlier, as presented in Eq.~\ref{eq:iqr_outliers}. These outliers represent the potential correct subkey candidates that do not follow the distribution of the wrong subkeys. The incorporation of the IQR metric enhances the evaluation process by facilitating the identification of the correct subkey candidate. This, in turn, allows for a more accurate assessment of side-channel vulnerabilities.

\section{Experimental Setup}
\label{sec:experimental_setup} \noindent
In this section, we describe the hardware setup and data collection steps for each power and impedance side channel.

\subsection{Hardware Setup} 
\begin{figure}[!htbp]
\centering
\includegraphics[width=0.45\textwidth]{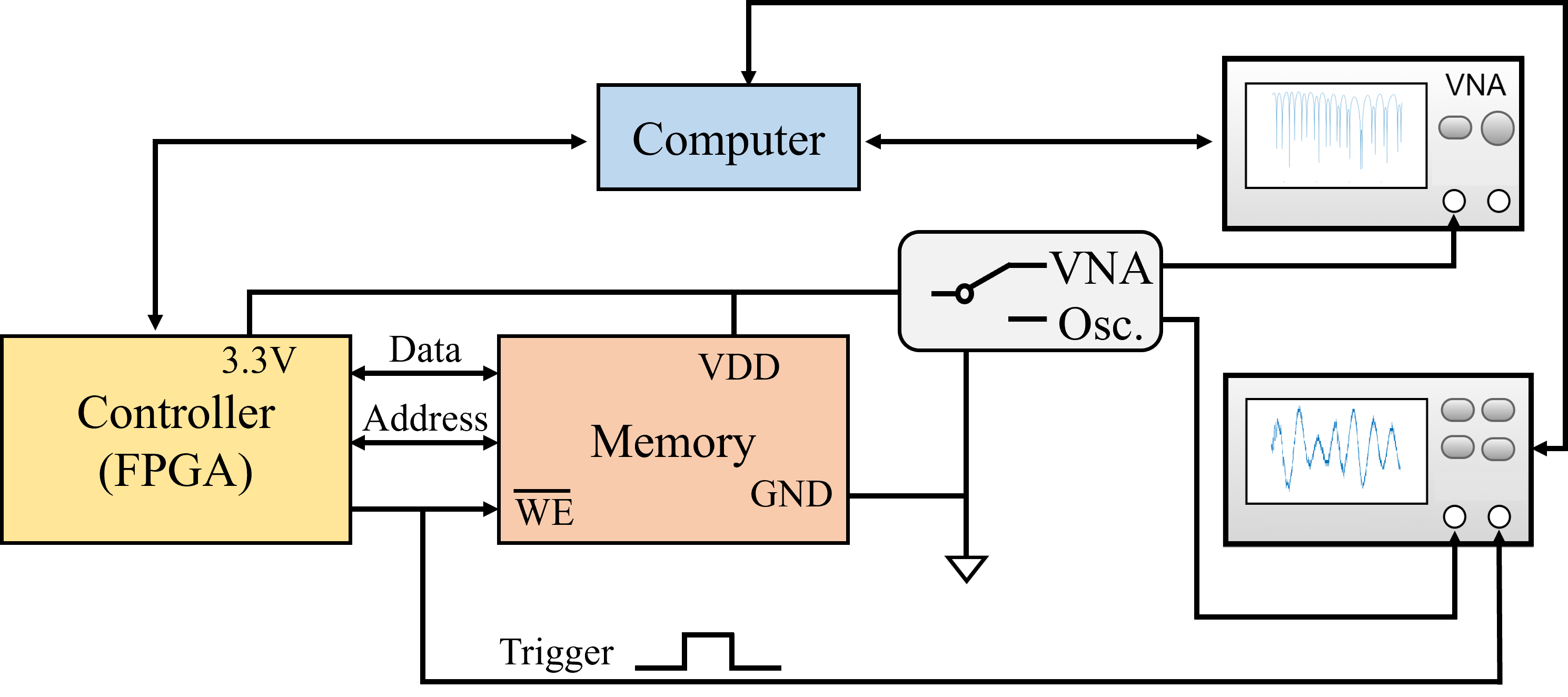}
\caption{Experimental setup diagram.}
\label{fig:experimental_setup}
\end{figure}

\noindent
The hardware setup is illustrated in Fig.~\ref{fig:experimental_setup}. We use an oscilloscope and a vector network analyzer (VNA) to measure the power and impedance signals, respectively. For power measurements, we employ the SIGLENT SDS5104X oscilloscope with a 5 Giga sample per second rate, enabling us to collect 2501 voltage sample points for each measurement. We utilize the Rigol RSA5032N spectrum analyzer for impedance measurements in VNA mode. We select a frequency range from 100 kHz to 3.2 GHz, allowing us to measure impedance at 10,001 linearly distributed frequency points.  
Additionally, we use the Alchitry Au development board with an Artix 7 FPGA~\cite{alchitry} to develop a memory controller. The memory controller controls the memory read and write operations. The Alchitry Au operates at a clock frequency of 100 MHz. 


\textbf{Power Side-Channel Attack:} 
During the measurement, we connect a probe of the oscilloscope to the 3.3V PDN of the memory chip. We use different plaintext to perform AES encryption and record the voltage fluctuations. We achieve synchronization of our measurements through the write-enable signal of the memory chip. We use an 8-bit plaintext and an 8-bit AES subkey in each experiment to compute an intermediate value. We record the voltage fluctuations at the time of writing the intermediate value of the first round of AES to the memory. This enables us to collect the necessary traces to extract all 16 subkeys of the 128-bit AES encryption.  

\textbf{Impedance Side-Channel Attack:}
For the impedance SCA, we connect the VNA to the 3.3V PDN of the memory chip via a coaxial cable. Unlike in the power SCA, we collect impedance measurements after the intermediate values have been written to the memory. The VNA measures the reflection coefficient parameters ($S_{11}$), which we convert to impedance values using Eq.~\ref{eq:S_to_Z}. Although impedance includes both phase and magnitude, we focus only on the magnitude in our analysis as it carries the most information~\cite{awal2022nearfield}.

\begin{figure}[!htbp]
\centering
\includegraphics[width=0.45\textwidth]{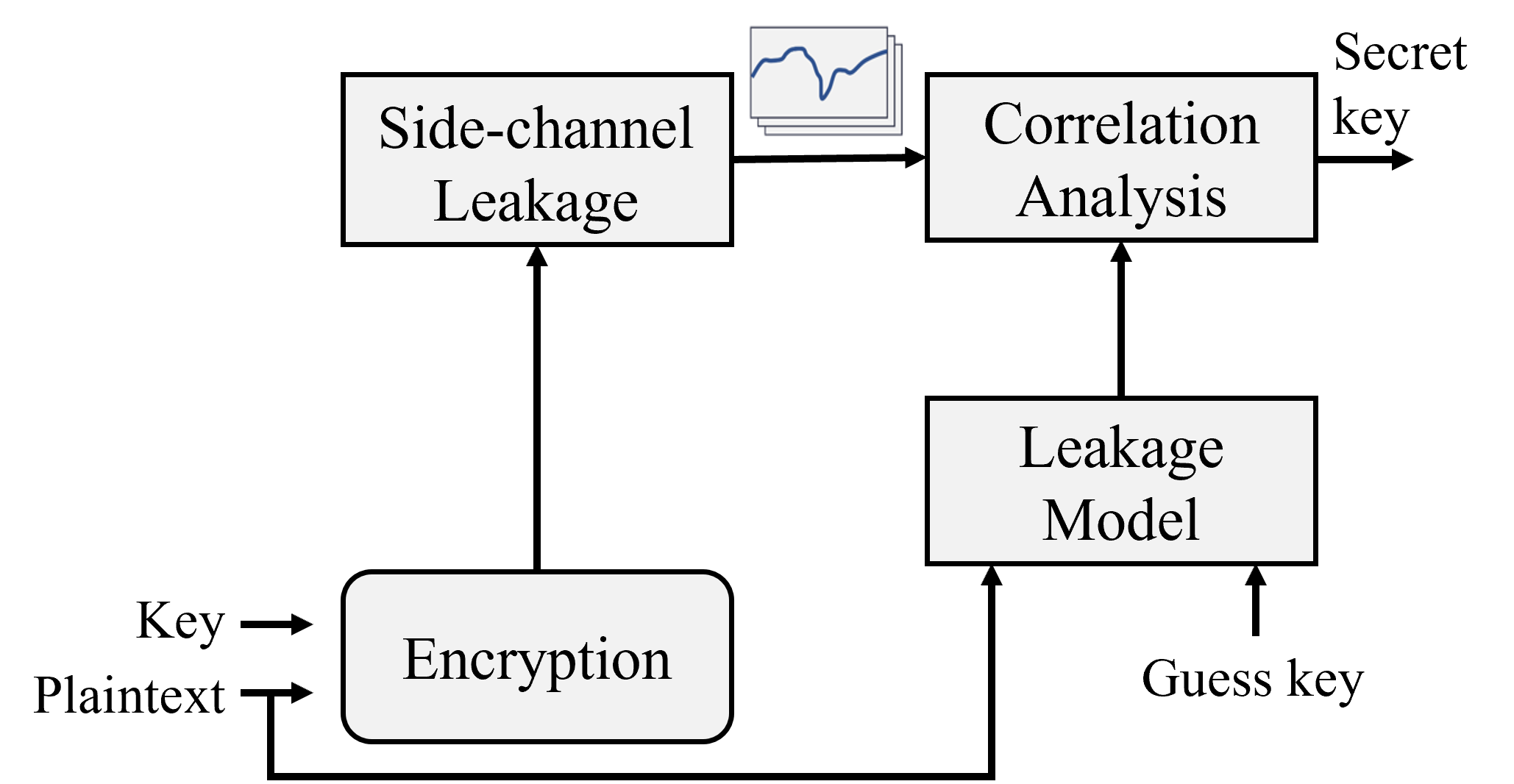}
\caption{AES key extraction method. 
}
\label{fig:attack_steps}
\end{figure}

\subsection{Data Collection and Analysis} \noindent 
For power side-channel measurements, we automate the data collection process by using  the \textit{write-enable} signal of the memory as a trigger. This trigger signal initiates the oscilloscope measurement. We develop a custom Python script incorporating standard commands for programmable instruments (SCPI) commands to transfer the measured traces to a computer. To mitigate noise, we take each measurement trace as an average of 128 measurements. These averaged traces are then used for the power SCA.

During the impedance side-channel measurements, we implement a similar automated setup. Python scripts with SCPI commands facilitate the transfer of measured traces from the VNA to the computer. Similar to the power side-channel measurements, we take the average of 100 measurements as each measurement trace to reduce noise. 

We perform correlation analysis with the collected measurement traces by comparing the actual leakages with hypothetical leakages. Fig.~\ref{fig:attack_steps} illustrates the analysis steps. We recover the correct key as the key value corresponding to the maximum correlation coefficient observed during the analysis.

\begin{figure*}[hbp]
\centering
\begin{subfigure}{.49\textwidth}
    \raggedright
    \includegraphics[width = 1.1\textwidth]{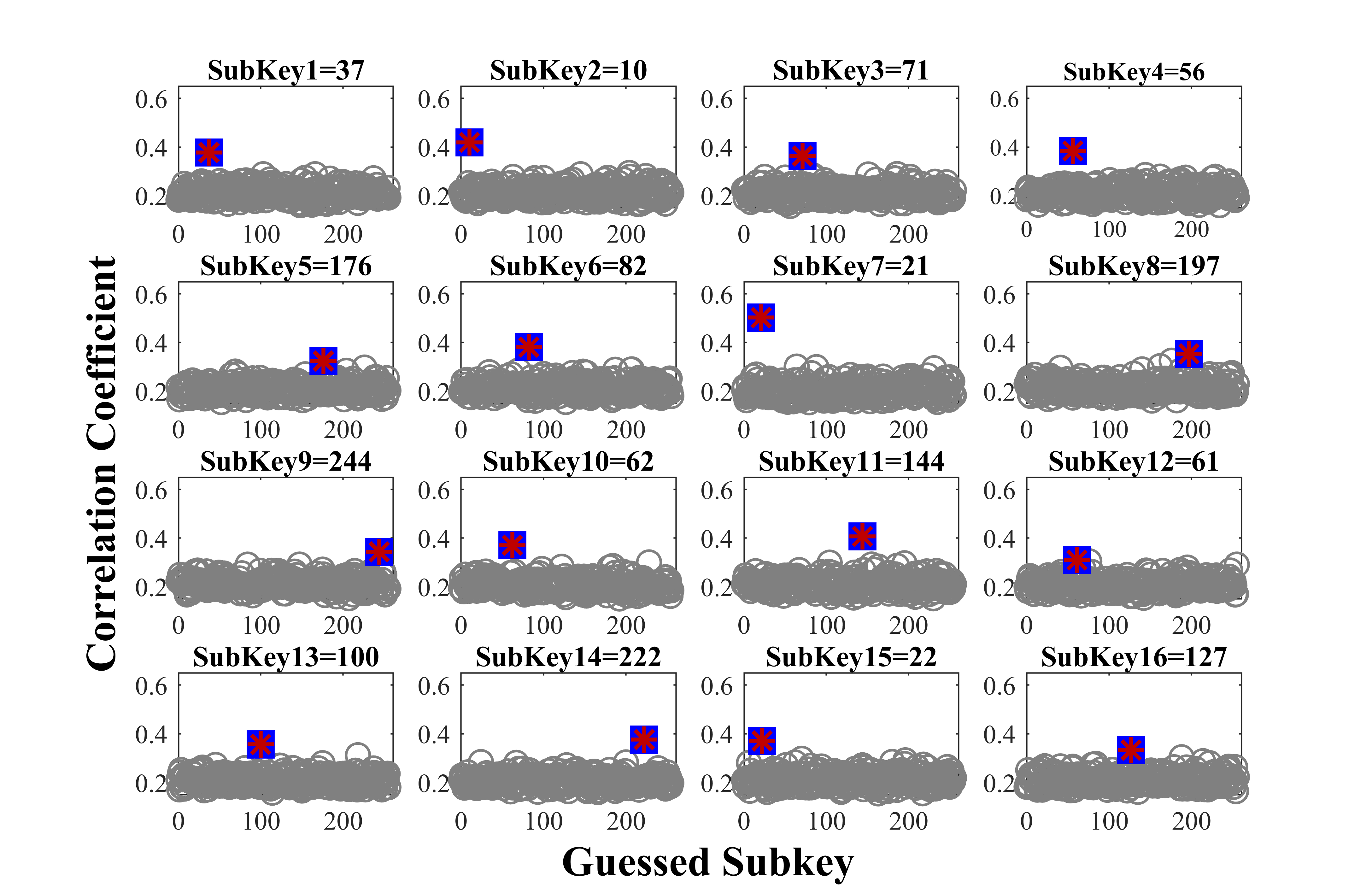}
    \caption{Power side-channel}
    \label{fig:power1}
\end{subfigure}
\begin{subfigure}{.49\textwidth}
    \raggedleft
    \includegraphics[width = 1.1\textwidth]{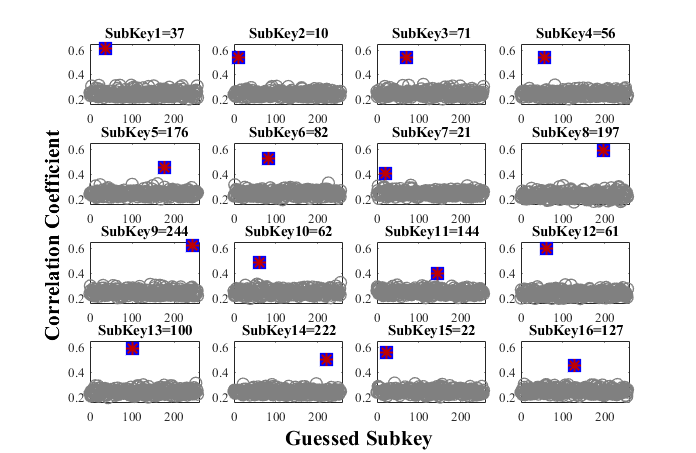}
    \caption{Impedance side-channel}
    \label{fig:impedance1}
\end{subfigure}
\caption{Results of 8-bit subkey extraction for 128-bit AES.}
\label{fig:corr_results_normal}
\end{figure*} 

\section{Results and Analysis: Power vs. Impedance}
\label{sec:results} \noindent
We use the AES-128 encryption and follow the evaluation metrics described in Section~\ref{sec:metric} to conduct a comprehensive comparison of the performance of power and impedance side channels. Additionally, in a subsequent scenario, we evaluate the resilience of these side channels in a noise-injected environment. In every case, we target the S-box substitution process during the first round of AES encryption. We adopt a divide-and-conquer approach by dividing the full 128-bit AES key into 16 subkeys and focus our efforts on recovering each subkey individually.

\subsection{Baseline Side-channel Analysis} \noindent
To initiate our analysis, we perform the correlation analysis described in Section~\ref{sec:correlation} for all possible 8-bit guessed subkeys and plot the findings in Fig.~\ref{fig:corr_results_normal}. Since we use the 128-bit AES encryption, our analysis involves 16 subkeys. 
Maximum correlation values in Fig.~\ref{fig:corr_results_normal} are highlighted by an asterisk ($*$), and the correlation corresponding to the correct subkey is distinguished by a square ($\square$) for visual clarity. Fig.~\ref{fig:power1} represents the results obtained from the power SCA, while Fig.~\ref{fig:impedance1} depicts the results of the impedance SCA. 
The results demonstrate the vulnerability of the 128-bit AES implementation to both the power and impedance side-channel attacks.

\begin{table}[!htbp]
    \centering
    \begin{tabular}{m{3.9em}|m{1.5em}m{1.5em}m{1.5em}m{1.5em}m{1.5em}m{1.5em}m{1.5em}m{1.5em}}
        \hline \hline 
         \textbf{Subkey} & \textbf{0x25} & \textbf{0x0A} & \textbf{0x47} & \textbf{0x38} & \textbf{0xB0} & \textbf{0x52} & \textbf{0x15} & \textbf{0xC5}\\
         \hline \\
         Power & 0.37 & 0.41 & 0.36 & 0.38 & 0.32 & 0.38 & 0.50 & 0.35\\\\
         Impedance & 0.61 & 0.54 & 0.54 & 0.53 & 0.45 & 0.52 & 0.40 & 0.58\\ \\
         \hline \hline
         \textbf{Subkey} & \textbf{0xF4} & \textbf{0x3E} & \textbf{0x90} & \textbf{0x3D} & \textbf{0x64} & \textbf{0xDE} & \textbf{0x16} & \textbf{0x7F}\\
         \hline \\
         Power & 0.34 & 0.37 & 0.40 & 0.30 & 0.35 & 0.37 & 0.37 & 0.33\\\\
         Impedance & 0.61 & 0.48 & 0.39 & 0.59 & 0.59 & 0.50 & 0.55 & 0.44 \\ \\ \hline
    \end{tabular}
    \caption{Maximum correlation coefficients observed.}
    \label{tab:max_correaltions}
\end{table}

\begin{figure}[!htbp]
\centering
\includegraphics[width=3.6in]{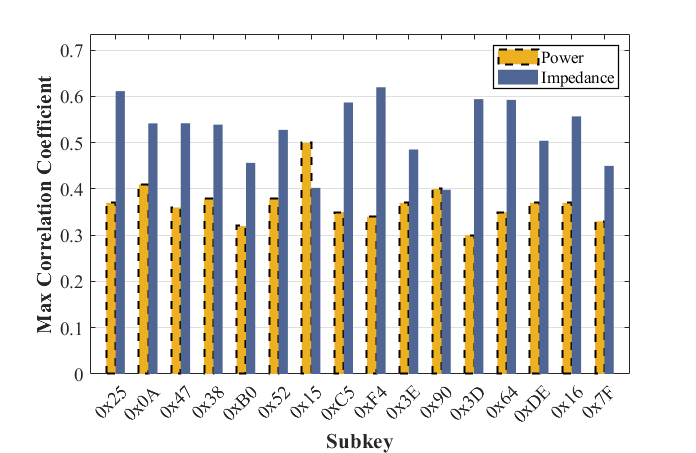}
\caption{Maximum correlation coefficient without noise.}
\label{fig:max_corr}
\end{figure}

Table~\ref{tab:max_correaltions} provides a summary of the maximum correlation coefficient observed in both side channels for each individual subkey. 
Additionally, we compare the maximum correlation coefficient values obtained in Fig.~\ref{fig:corr_results_normal} for each subkey. Fig.~\ref{fig:max_corr} presents a visual representation of Table~\ref{tab:max_correaltions}. The findings show that the maximum correlation values obtained through impedance side channels are consistently higher than those of the power side channels. This suggests a potential superiority of the impedance side channel in information leakage over the power side channel.

\begin{figure}[!htbp]
\centering
\includegraphics[width=3.6in]{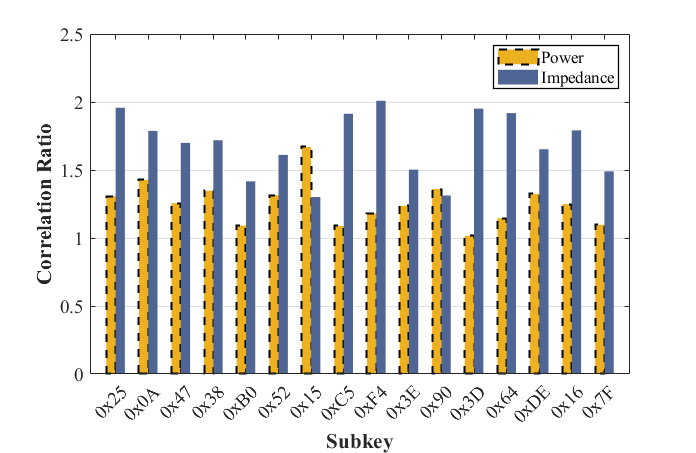}
\caption{Correlation ratio without noise.}
\label{fig:corr_ratio}
\end{figure}

\textbf{Correlation Ratio:} We proceed to compare and analyze the results obtained from the two side channels using the metrics discussed in Section~\ref{sec:metric}. First, to assess the discriminability level of the recovered correct key concerning other key candidates, we calculate the correlation ratio for each subkey using Eq.~\ref{eq:corr_ratio_modified}. The resulting graph is presented in Fig.~\ref{fig:corr_ratio}. The findings further suggest that the correct subkeys can be identified with a higher discriminability level most of the time when using the impedance side channel compared to the power side channel. 

\textbf{MTD:} As the second evaluation metric, we use the MTD in recovering the 128-bit AES key. The MTD represents the minimum number of side-channel traces required to successfully recover the secret key or a part of it, such as a subkey. 
When comparing the performance of the two side channels for AES-128 subkey extraction, the MTD metric aids in determining which side channel is more effective. The side channel with a lower MTD value is considered more potent, as it can recover the subkey using fewer traces, posing a higher security threat to the cryptographic system. 

\begin{figure}[!htbp]
\centering
\begin{subfigure}{.49\textwidth}
    \raggedleft
    \includegraphics[width = 1\textwidth]{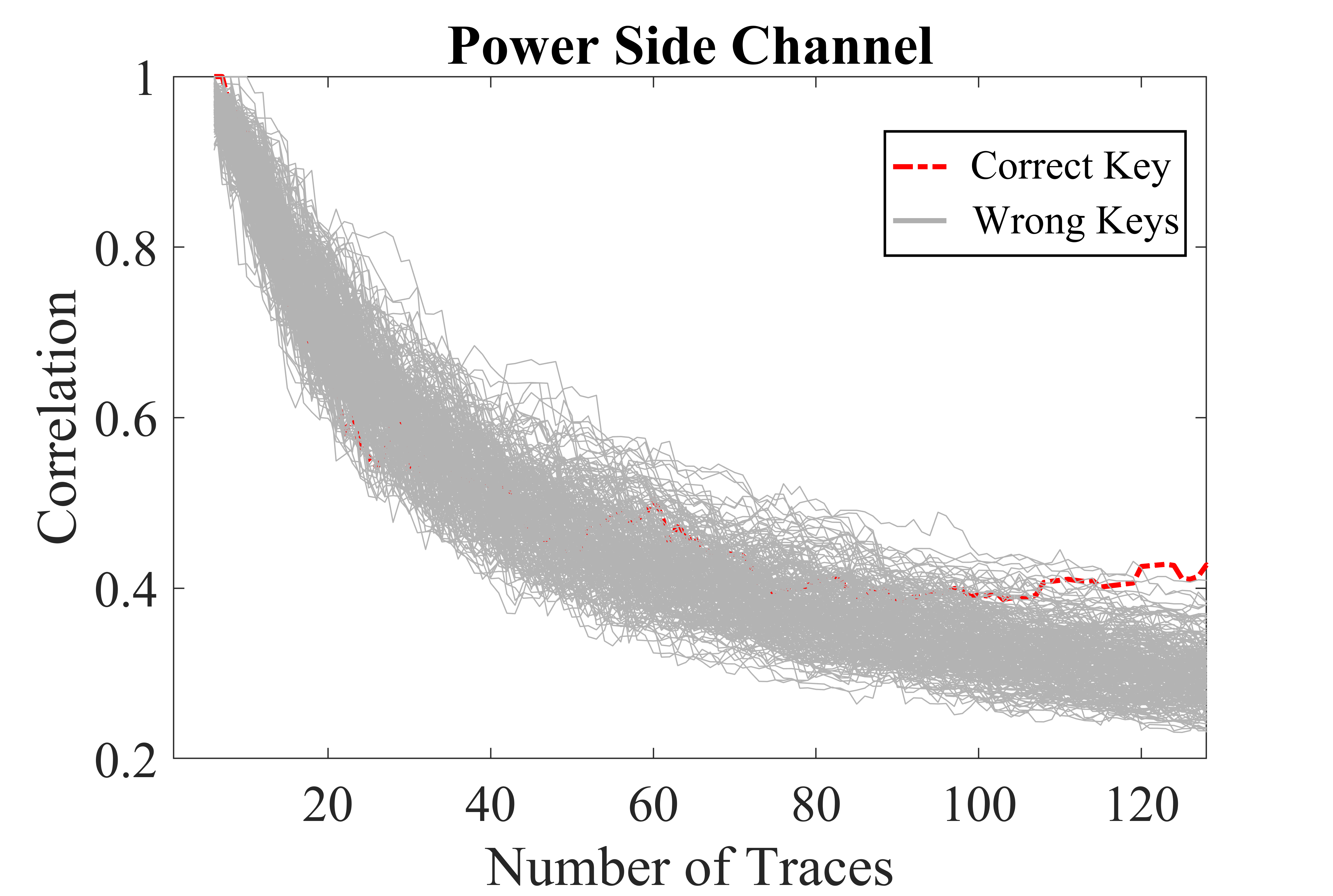}
    \caption{Power side-channel}
    \label{fig:MTD_power}
\end{subfigure}
\begin{subfigure}{.49\textwidth}
    \raggedright
    \includegraphics[width = 1\textwidth]{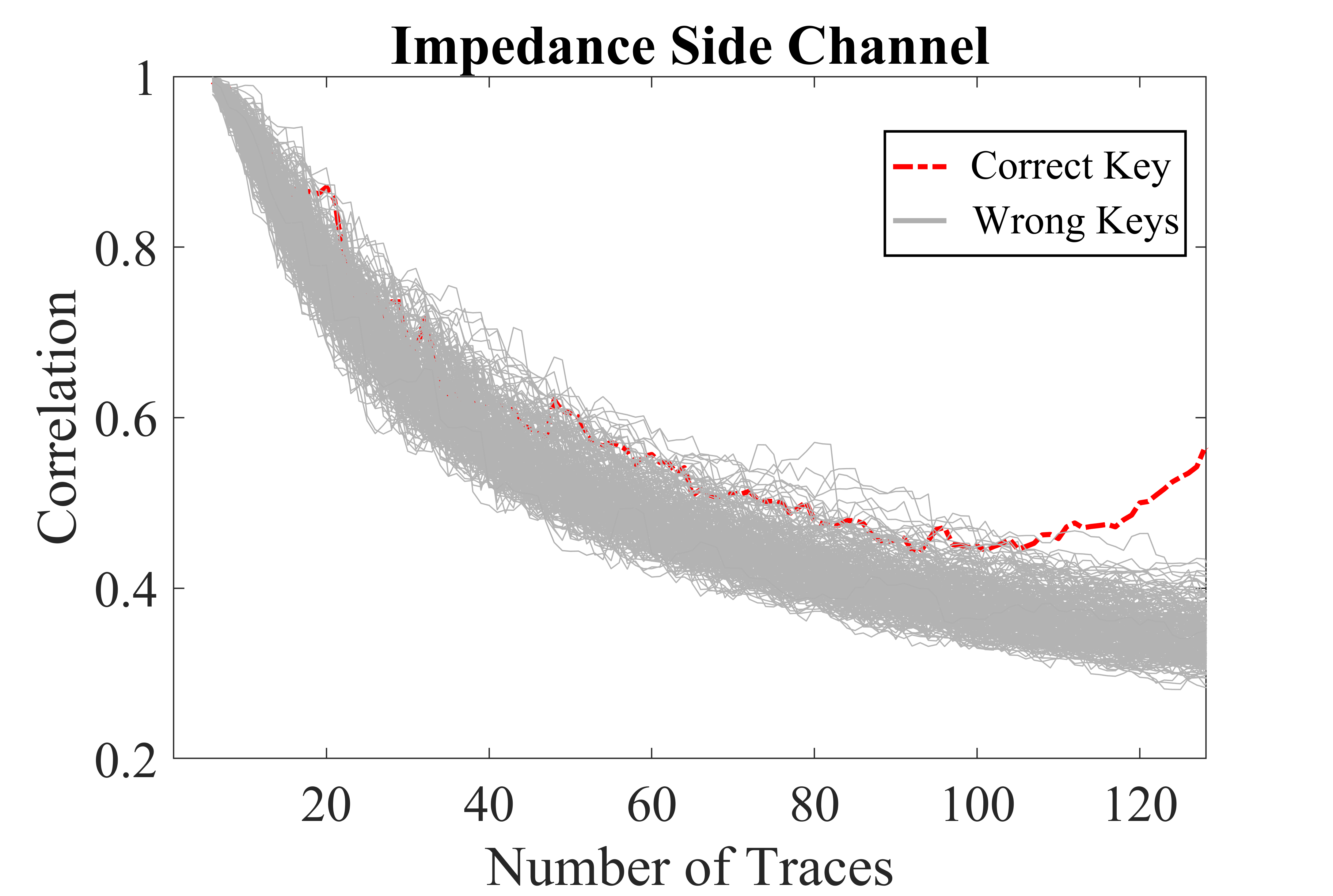}
    \caption{Impedance side-channel}
    \label{fig:MTD_impedance}
\end{subfigure}
\caption{Subkey-1 recovery with 128 traces. 
}
\label{fig:MTD}
\end{figure}

Our observations indicate that the MTD performs better in the impedance side-channel than the power side-channel. We present the MTD for subkey-1 in Fig.~\ref{fig:MTD} for only 128 traces. We find that with this limited number of traces, the correlation coefficient of the correct key is distinctly identified from the wrong key guesses using the impedance side-channel. In contrast, in the case of the power side-channel, the correlation coefficient of the correct key falls within the range of correlation coefficients of the wrong key guesses. We observe similar findings for other subkeys as well. The results suggest that the correct key can be recovered with fewer measurements using the impedance side channel compared to the power side channel. 

\begin{table}[!htbp]
    \centering
    \begin{tabular}{m{3.9em}|m{1.5em}m{1.5em}m{1.5em}m{1.5em}m{1.5em}m{1.5em}m{1.5em}m{1.5em}}
        \hline \hline 
         \textbf{Subkey} & \textbf{0x25} & \textbf{0x0A} & \textbf{0x47} & \textbf{0x38} & \textbf{0xB0} & \textbf{0x52} & \textbf{0x15} & \textbf{0xC5}\\
         \hline \\
         Power & 1 & 1 & 1 & 1 & F & 1 & 1 & F \\\\
         Impedance & 1 & 1 & 1 & 1 & 1 & 1 & 1 & 1 \\ \\
         \hline \hline
         \textbf{Subkey} & \textbf{0xF4} & \textbf{0x3E} & \textbf{0x90} & \textbf{0x3D} & \textbf{0x64} & \textbf{0xDE} & \textbf{0x16} & \textbf{0x7F}\\
         \hline \\
         Power & F & F & 1 & F & F & 1 & 1 & F \\\\
         Impedance & 1 & 1 & 1 & 1 & 1 & 1 & 1 & 1 \\ \\ \hline
    \end{tabular}
    \caption{Number of outliers using the highest five correlation coefficients. 
    }
    \label{tab:IQR_confidence}
\end{table}

\textbf{IQR:} To better understand the results and quantify the confidence in the recovered keys, we employ the IQR method on the highest five correlation coefficients obtained from Fig.~\ref{fig:corr_results_normal}. This approach helps us identify the guessed key for which the corresponding correlation coefficient is an outlier. Such an outlier indicates that the correlation coefficient can be statistically distinguished due to that guessed key, making it a potential correct subkey candidate. In scenarios where we find multiple outliers, all of them are considered potential correct subkey candidates. However, the absence of outliers implies that we cannot reliably identify the correct key based on the correlation coefficient values, indicating poor performance.

Table~\ref{tab:IQR_confidence} presents the correlation coefficients that lie outside the first and third quartile range. The `F' in the table signifies a failure to distinguish the correct subkey from the wrong ones. The findings suggest that the impedance side channel provides more distinct outliers compared to the power side channel, further corroborating the superiority of the impedance side channel in extracting the correct subkeys.


\subsection{Noise-Injected Side-channel Analysis}  \noindent
In the second phase of the experiment, we introduce noise into the measurement traces by executing background activities in the memory controller while collecting power and impedance measurements. To implement these background activities, we use ten 5-bit linear feedback shift registers (LFSRs). Each of these LFSRs is designed to shift bits to the left. 
Each LFSR operates by shifting bits to the left and uses XOR logic 
to generate an input bit at each clock cycle. In our implementation, we tap the fourth and fifth bits of the LFSR to generate the new input bit. The use of LFSRs ensures the generation of a sequence of bits that appears random. Moreover, the parallel use of ten such left-shift LFSRs allows us to generate the background activities efficiently. 

\begin{figure*}[!htbp]
\centering
\begin{subfigure}{.49\textwidth}
    \raggedright
    \includegraphics[width = 1.1\textwidth]{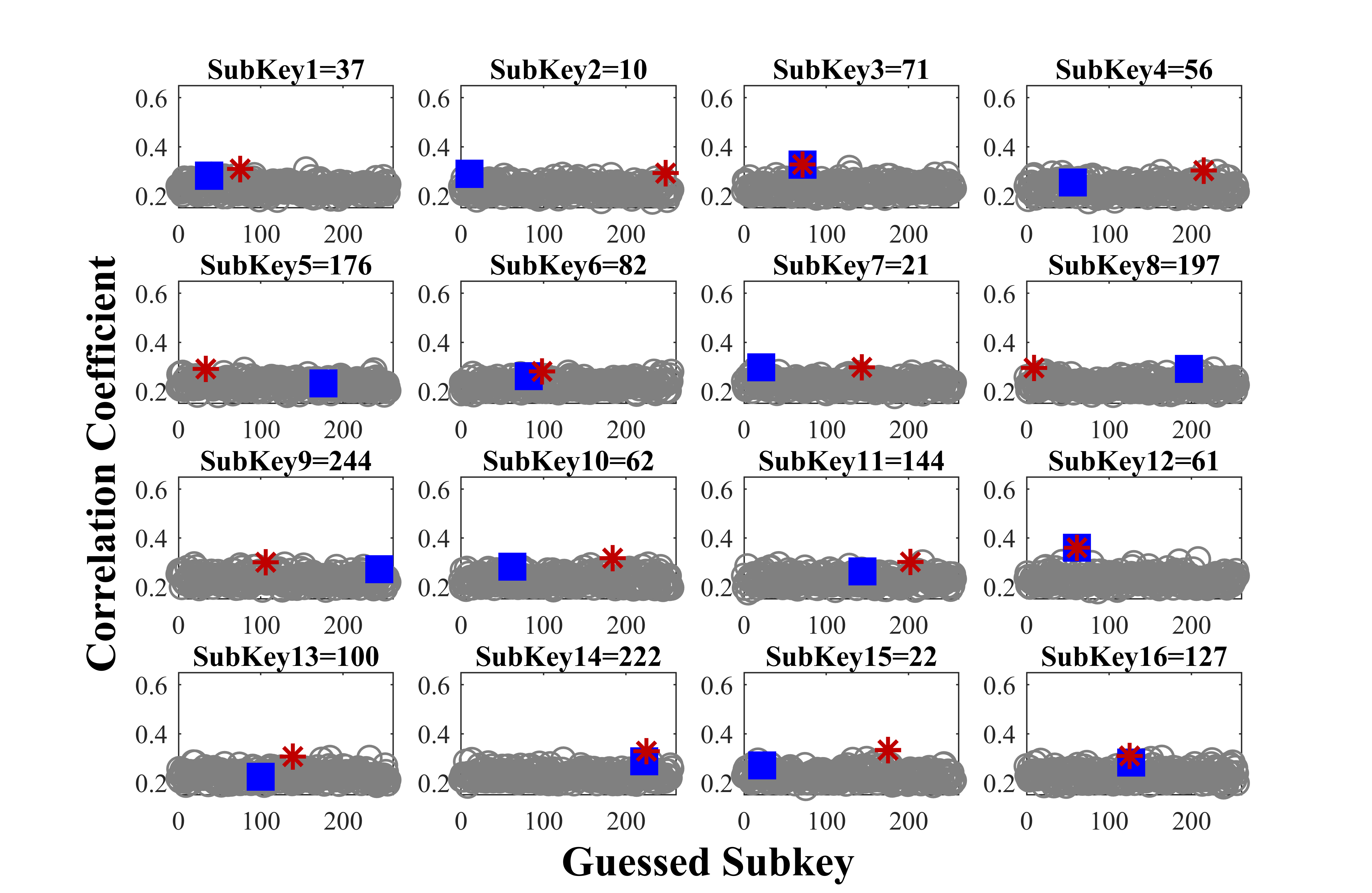}
    \caption{Power side-channel}
    \label{fig:power2}
\end{subfigure}
\begin{subfigure}{.49\textwidth}
    \raggedleft
    \includegraphics[width = 1.1\textwidth]{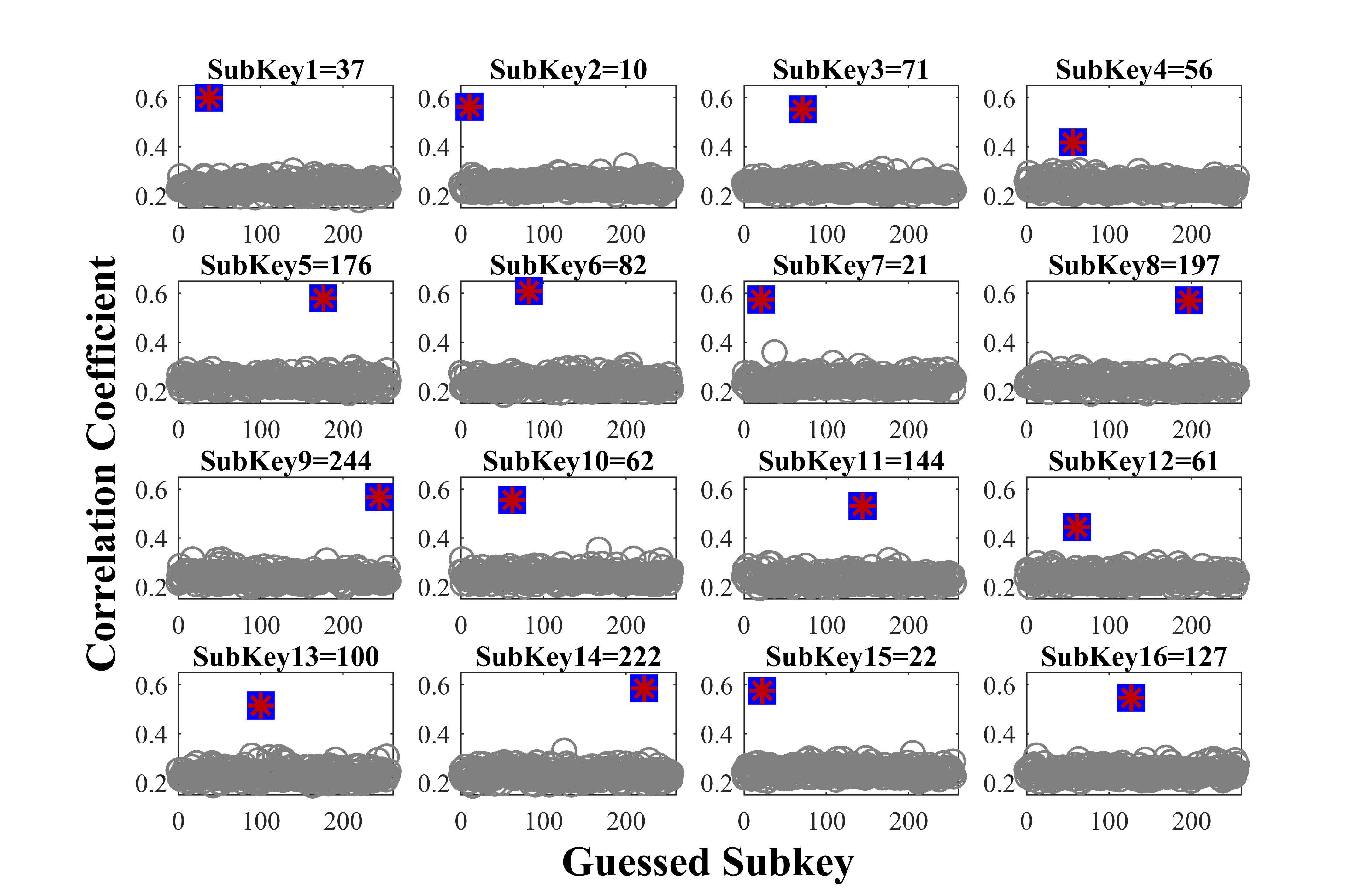}
    \caption{Impedance side-channel}
    \label{fig:impedance2}
\end{subfigure}
\caption{Results of 8-bit subkey extraction fro 128-bit AES with noise.}
\label{fig:corr_results_lfsr}
\end{figure*}

Following the LFSR implementation, we collect the same number of traces as previously for each power and impedance side channel setup and repeat the correlation analysis for AES-128 key extraction. Fig.~\ref{fig:corr_results_lfsr} presents the subkey extraction results of each side channel with the LFSR implemented in the background. As illustrated, the performance of the power side channel in extracting the subkeys degrades significantly in the presence of noise. 
However, even under noisy conditions, the impedance side channel can successfully extract all the 8-bit subkeys of AES-128.

Considering the results of the second attack scenario after introducing additional noise to the measurements, we find that the impedance side channel can still recover all the subkeys correctly, in contrast to the power measurements. These background activities result in power consumption variations due to increased internal switching activities. 

\subsection{Summary of Results} \noindent
This comparative analysis concludes several key insights regarding the efficacy and implications of impedance and power side-channel attacks within cryptographic systems: 

\begin{itemize}
    \item Impedance SCA demonstrates superior efficacy in key extraction, demonstrating its potential as a significant security and privacy threat that may surpass that of traditional power leakage attacks, both in noisy and noise-free environments. This capability underscores the importance of addressing impedance-based vulnerabilities in cryptographic systems. 
    \item  Some countermeasures that can mitigate power side-channel attacks (e.g., background noise injection) may not provide equivalent protection against impedance side-channel attacks. However,  further case studies and analyses are required to generalize this statement effectively. By exploring a diverse range of scenarios and configurations, we can better understand the specific conditions under which impedance side-channel attacks may succeed or be thwarted. This comprehensive approach will enable us to identify patterns and establish more definitive conclusions regarding the limitations of existing countermeasures against these emerging vulnerabilities. 
\end{itemize}

\section{Discussion on the Comparison} \noindent
 In our comparative analysis between impedance-based SCA and power-based SCA, we observed that impedance SCA consistently outperforms power SCA in terms of signal strength and data extraction accuracy. One of the key advantages of impedance SCA is its ability to precisely control the device's clock, allowing the clock to be paused at specific intervals during impedance measurement. This enables the isolation of switching noise that typically arises from unrelated background activities, ensuring that the observed signal is free from interference. In contrast, power SCA relies on measuring dynamic power consumption, which primarily tracks the flow of current during operation. This method captures not only the switching activity of the desired operation but also the background noise generated by other concurrent switching events within the circuit. The inability to filter out this extraneous noise results in a less distinct signal, making power SCA more susceptible to inaccuracies and noise contamination.

Additionally, impedance measurements are more sensitive to subtle variations in the physical properties of the device, such as material characteristics or component wear, which can provide more nuanced insights into the device’s behavior. Power consumption, by comparison, offers a broader and less differentiated view of the device's operation, making it harder to isolate specific activities or transitions. Additionally, impedance SCA benefits from lower bandwidth requirements compared to power SCA, as the impedance measurement process does not need to capture the fast transient currents that power SCA requires. This reduced bandwidth requirement simplifies the data acquisition process and enhances the accuracy of the collected data by limiting high-frequency noise. 


Importantly, impedance side-channel vulnerabilities remain exploitable even when the chip is powered off, particularly in the context of non-volatile memory. In such cases, the impedance characteristics associated with stored data differ and can be measured without the need to power up the device. This occurs because the impedance of a chip, influenced by specific stored data, does not rely on the device's power state or power consumption characteristics. This presents a significant challenge for traditional power-based countermeasures, such as power gating. 
While power gating effectively mitigates power SCA risks by shutting down power to specific sections of a chip during idle periods, it fails to address the vulnerabilities inherent in impedance measurements. As a result, a more comprehensive security approach is necessary.

\section{Conclusion}
\label{sec:conclusion} \noindent
This study provides a comprehensive comparative analysis of impedance and power SCA techniques, focusing specifically on their efficacy in extracting cryptographic keys. Contrary to the common perception that views impedance SCA merely as an alternative to power analysis, our findings find that impedance analysis holds better potential in extracting cryptographic   keys, including scenarios where power analysis proves inadequate. 

The findings demonstrate the robustness and effectiveness of impedance analysis, challenging its often-overlooked significance. Additionally, the results of this study serve as a compelling motivation for further exploration into the underlying mechanisms of impedance analysis and their implications for hardware security. Furthermore, our findings underscore the need to incorporate impedance SCA into the standard evaluation protocols for cryptographic systems. By doing so, we can ensure a more comprehensive and robust assessment of potential vulnerabilities, thereby enhancing the overall security posture. 




\normalem
\bibliographystyle{IEEEtran}
\bibliography{IEEEabrv,references}

\begin{IEEEbiography}[{\includegraphics[width=1in,height=1.25in,clip,keepaspectratio]{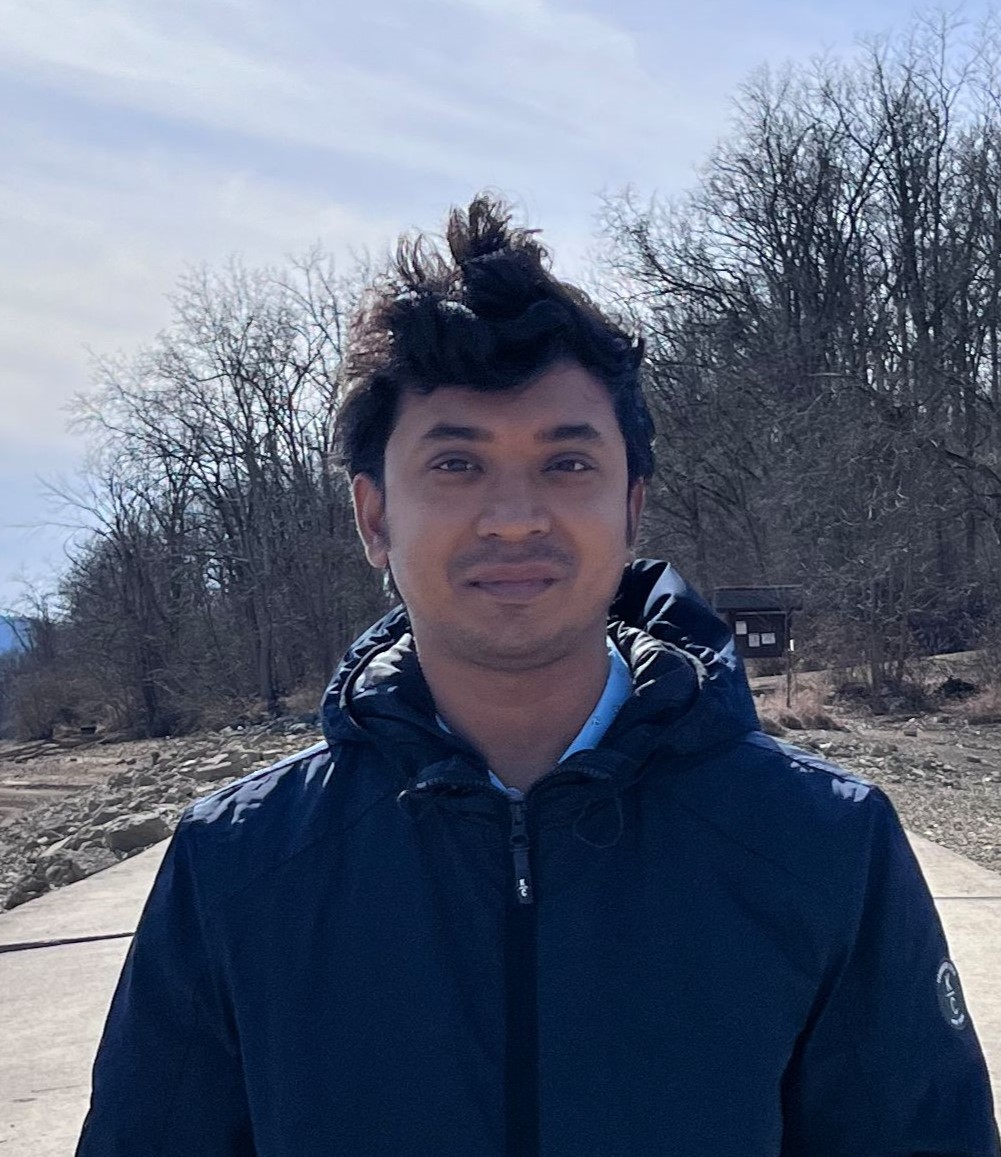}}]{Md Sadik Awal} (S'17) received his B.Sc. degree in Electrical and Electronic Engineering from Bangladesh University of Engineering and Technology, Bangladesh in 2021. Since 2022, he has been a Graduate Research Assistant in the SeRLoP Lab while pursuing a Ph.D. at Florida International University's School of Electrical and Computer Engineering. His current research interests span the areas of hardware security, side-channel analysis, embedded systems, and signal processing.
\end{IEEEbiography}
\vspace{-1.2cm}
\vskip 0pt plus -1fil 

\begin{IEEEbiography}[{\includegraphics[width=1in,height=1.25in,clip,keepaspectratio]{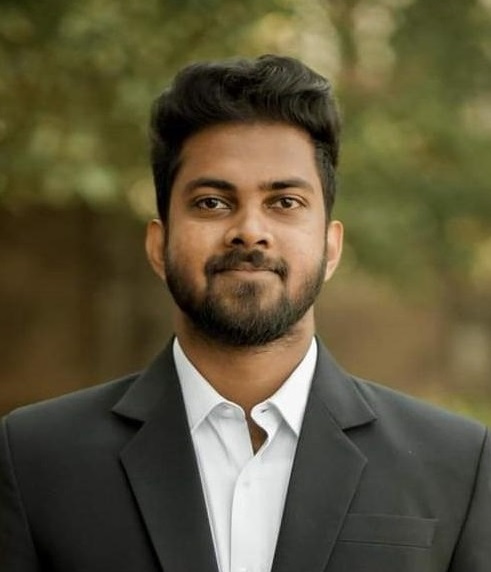}}]{Buddhipriya Gayanath}
is a graduate student at the Department of Electrical and Computer Engineering at Florida International University. He earned his B.Sc. degree from University of Sri Jayewardenepura in 2022. 
His current research focuses on radio frequency, analog and mixed-signal circuits, and signal processing.
\end{IEEEbiography}
\vspace{-1.2cm}
\vskip 0pt plus -1fil 

\begin{IEEEbiography}[{\includegraphics[width=1in,height=1.25in,clip,keepaspectratio]{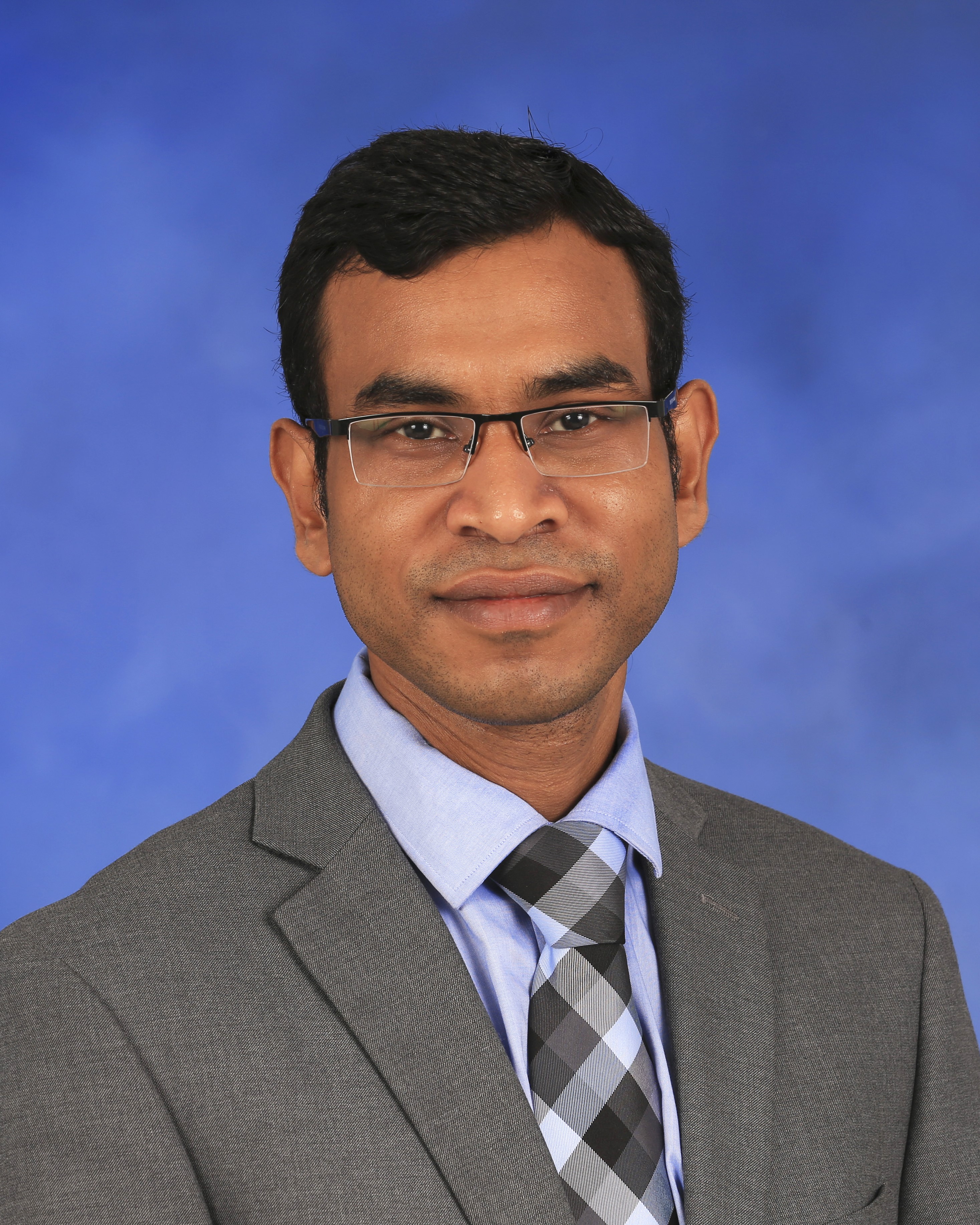}}]{Md Tauhidur Rahman} (S'12–M'18-SM'21) is an Assistant Professor in the Department of Electrical and Computer Engineering at Florida International University (FIU). He received his Ph.D. degree in Computer Engineering from the University of Florida in 2017. His current research interests include hardware security and trust, side-channel analysis, memory systems, embedded security, and privacy. He is one of the recipients of the 2019 NSF CRII Award. 
\end{IEEEbiography}

\vfill

\end{document}